\documentclass{aa}  

\usepackage{graphicx}
\usepackage{adjustbox}
\usepackage{multirow}
\usepackage{tablefootnote}
\usepackage[rightcaption]{sidecap}
\usepackage{floatrow}
\usepackage{txfonts,textcomp}
\usepackage{hyperref}
\begin{document}

\newbox\grsign \setbox\grsign=\hbox{$>$} \newdimen\grdimen \grdimen=\ht\grsign
\newbox\simlessbox \newbox\simgreatbox
\setbox\simgreatbox=\hbox{\raise.5ex\hbox{$>$}\llap
     {\lower.5ex\hbox{$\sim$}}}\ht1=\grdimen\dp1=0pt
\setbox\simlessbox=\hbox{\raise.5ex\hbox{$<$}\llap
     {\lower.5ex\hbox{$\sim$}}}\ht2=\grdimen\dp2=0pt
\def\simgreat{\mathrel{\copy\simgreatbox}}
\def\simless{\mathrel{\copy\simlessbox}}
\newbox\simppropto
\setbox\simppropto=\hbox{\raise.5ex\hbox{$\sim$}\llap
     {\lower.5ex\hbox{$\propto$}}}\ht2=\grdimen\dp2=0pt
\def\simpropto{\mathrel{\copy\simppropto}}

\title{Abundances in eight bulge stars from the optical and near-infrared}

 \author{Patr\'{\i}cia da Silva \inst{1,2} 
 \and B. Barbuy \inst{2} \and H. Ernandes \inst{3} \and S. O. Souza \inst{2} \and J. G. Fern\'andez-Trincado \inst{4} \and D. Gonz\'alez-D\'iaz \inst{4} \thanks{Observations collected at the European Southern Observatory, Paranal, Chile (ESO),
 under programmes 65.L-0340A, 93.D-0123A 97.D-0175A, 0103.D-0828A, 71.B--517 and 73.B-0074; and APOGEE project. }  }

\institute{Observatoire de Paris, LERMA, Coll\`ege de France, CNRS, PSL University, Sorbonne University, 75014, Paris  -- e-mail: patricia2.silva@alumni.usp.br 
\and 
Instituto de Astronomia, Geof\'isica e Ci\^encias Atmosf\'ericas, Departamento de Astronomia, Universidade de S\~ao Paulo, 05508-090, SP, Brazil
\and 
Lund Observatory, Department of Astronomy and Theoretical Physics, Lund University, Box 43, SE-221 00 Lund, Sweden 
\and 
Instituto de Astronom\'ia, Universidad Cat\'olica del Norte, Av. Angamos 0610, Antofagasta, Chile
}
            
   \date{Received ....; accepted .....}

 
  \abstract
 {The abundances of the $\alpha$-elements are key for understanding the early chemical enrichment of the Galactic bulge. The elements of interest present lines in different wavelength regions, and some of them show lines only in part of the spectra. In the present work, the CNO trio, the $\alpha$-elements Mg, Si, Ca, and Ti, and odd-Z Na and Al are examined as measured from optical and \textit{H}-band lines. }
   {The aim of this work is to carry out a detailed comparison of stellar parameters and abundances 
   derived in the optical and near-infrared (\textit{H}-band).
   We also inspect the best available lines for a list of 
   bulge stars previously analyzed by the Apache Point Observatory Galactic Evolution Experiment (APOGEE) team in the H-band and by our group in the optical.
   This work is mainly of interest to spectroscopists.}
   {In the present work, we compared the stellar parameters and abundance results derived from  APOGEE \textit{H}-band spectra with optical analyses based on Ultraviolet and Visual Echelle Spectrograph at the Very Large Telescope (VLT/UVES)
   data for eight bulge stars. } 
   {We point out the most suitable wavelength region for each of the studied elements, and highlight difficulties in the
   derivation of stellar parameters both in the optical and \textit{H}-band. The near-infrared will allow observations of a large number of stars in the near future given new instruments soon to be available. The identification of spectral  lines in this spectral region and the investigation of their 
reliability are ongoing efforts worldwide. New instruments will also allow simultaneous observation of \textit{H}-band and optical.
   }
   {}


   \maketitle


\section{Introduction}

Abundances in the Galactic bulge are being scrutinized for a better definition of
the fast star formation rate and identification of the supernovae that enriched the
medium very early on (e.g., \citealt{{barbuy18a},matteuci21}). This applies to field and globular cluster stars, with the latter
also involving new progress in our understanding of self-enrichment and the multiple
stellar populations that undoubtedly occurred in almost all clusters \citep{piotto15,renzini15,milone17,bastian18,milone22}.

The light elements are discriminators of the type of supernova and nucleosynthesis that formed them, and in order to use them, reliable abundance values are needed. Consequently, in this work, we investigate the possible differences between the abundance derivation from the optical and the near-infrared (NIR) for eight stars analyzed in the optical by our group and also available from NIR spectra.

Instrumentation in astronomy allowing observations in the NIR has progressed significantly in the recent past. Apache Point Observatory Galactic Evolution Experiment (APOGEE--  \citealt{majewski17}) allows high-resolution spectroscopy in the \textit{H}-band at a resolution of R $\sim$ 22 000 carried out at the 2.5m Sloan Foundation Telescope at the Apache Point Observatory in New Mexico (APOGEE-1 and APOGEE-2N: \citealt{beaton21}), and the 2.5m du Pont Telescope at Las Campanas Observatory in Chile (APOGEE-2S). Spectroscopy in the JHK bands will be intensified with the James Webb Space Telescope (JWST), and new spectrographs on ground-based telescopes such as the Multi-Object Optical and Near-IR Spectrograph (MOONS) at the Very Large Telescope (VLT), the Warm INfrared Echelle spectrograph to REalize Decent high-resolution spectroscopy (WINERED) at the Magellan telescope (see \citet{minniti24}), and the Multi-Object Spectrograph for Astrophysics, Intergalactic-medium studies and Cosmology (MOSAIC) at the Extremely Large Telescope (ELT) at the European Southern Observatory (ESO), among others. 

The aim of the present work is to compare available lines and results that can be obtained in the optical and \textit{H}-band for a list
of elements. This study is useful as it indicates the most reliable lines in the optical and/or \textit{H}-band ---which is
of particular interest for observations with the ELT/MOSAIC that should allow simultaneous observations in the NIR and selected regions in the optical at resolutions of 6,000 and 20,000 \citep{hammer21}.

The analysis of \textit{H}-band spectra in the APOGEE project is carried out through a Nelder-Mead algorithm \citep{nelder65}, which simultaneously  fits the stellar parameters effective temperature (T$_{\rm eff}$, gravity (log~g), metallicity ([Fe/H]), and microturbulence velocity (v$_{\rm t}$), together with the abundances of carbon, nitrogen, and alpha-elements with the APOGEE Stellar Parameter and Chemical Abundances Pipeline - ASPCAP pipeline, which is based on the FERRE code \citep{garcia-perez16}. Spectroscopists  within the APOGEE community are carrying out continuous and intensive work, but  improvements
are still needed for some elements and lines.

\citet{jonsson18} carried out an important verification of the accuracy of the ASPCAP results, and reported previous analyses with the same intent: \citet{hawkins16} used the DR12 APOGEE to analyze stars with asteroseismic analysis from Kepler light curves, and found differences regarding Si, S, Ti, and V. \citet{souto16} used the DR13 APOGEE to analyze spectra of 12 giant stars belonging to the open cluster NGC 2420 ([Fe/H] $\sim$ -0.16), and found differences in the abundances of Na, Al, and V. \citet{fernandez-trincado16} manually reanalyzed the DR12 APOGEE spectrum of one peculiar metal-poor field giant star with a globular cluster second-generation abundance pattern, and found differences in the abundances of C, N, O, Mg, and Al of about 0.3 dex.

\citet{holtzman18} described the series of technical steps of the ASPCAP procedure, and the grids of synthetic spectra used to derive stellar parameters. \citet{jonsson18} carried out a comparison between results from DR13 and DR14, as well as selected literature results, with mean differences reported in their Table \ref{abundanceuves}.
\citet{jonsson20} and \citet{hayes22} studied the details of several lines from DR16 and DR17 results, and pointed out problems and limitations.

In \citet{razera22}, our group analyzed 58 moderately metal-poor bulge field  stars ($-$2.0 $<$ [Fe/H] $<$ $-$0.8) from APOGEE spectra, and identified the most suitable spectral lines for deriving the
abundances of C, N, O, Mg, Si, Ca, and Ce given that some of the ASPCAP adopted lines are too faint in these metal-poor stars. These authors confirmed the Mg, Si, and Ca from ASPCAP, whereas O and Ce were found to be  more abundant than the ASPCAP values.

In the present work, we identified eight stars previously analyzed  by our group from optical high-resolution spectra collected with the Ultraviolet and Visual Echelle Spectrograph (UVES) at the VLT. Importantly, these sources were also observed with APOGEE. For this sample, we carried out a detailed comparison of  stellar parameters and abundance results and inspected the available lines for a list of elements.

In Section \ref{sec2}, we list the sample stars and report previous calculations. In Section \ref{sec3}, we describe the present calculations.
In Section \ref{sec4} we compare the stellar parameters and abundances derived from the optical and near-infrared, and show how we carried out fits to lines both in the optical and in the near-infrared. We discuss our results in Section \ref{sec5} and draw conclusions in Section \ref{sec6}.

\section{The sample}\label{sec2}

The sample consists of eight stars analyzed in the optical using the UVES spectrograph at the VLT (ESO), and for which there are available observations with APOGEE. These stars are from the following globular clusters: one star from HP~1 \citep{barbuy16}, two stars from NGC~6558 \citep{barbuy18a}, one star from NGC~6522 \citep{barbuy14}, 
 and one star from Palomar~6 \citep{souza21}.  In \citet{ernandes2018} the abundances of Sc, V, Mn, Cu, and Zn were derived for most of these stars.
The sample is completed with three bulge field stars analyzed  for different elements, namely
C, N, and O in \citet{zoccali06,lecureur07,jonsson17}, and \citet{friaca17} (the latter is adopted),  Mn and Zn in \citet{barbuy13,barbuy15} and \citet{dasilveira18}, and Co and Cu in \citet{ernandes20}.

The \textit{H}-band data are from the APOGEE project, which is part of the Sloan Digital Sky Survey III and IV (SDSS; \citealt{blanton17}). The project aimed to investigate Milky Way stars observed at high resolution and high signal-to-noise ratios (S/Ns) in the \textit{H}-band.
The project SDSS-IV technical summary, the SDSS telescope, and APOGEE spectrograph are described in \citet{blanton17,gunn06,wilson19}, respectively, whereas \citet{zasowski13,zasowski17,beaton21,santana21} describe the APOGEE and APOGEE-2 target selections. The data release 17 (DR17) contains high-resolution (R $\sim$ 22,500 ) \textit{H}-band spectra (15140-16940 Å) for about 7$\times$10$^{5}$ stars. The 2.5m Sloan Foundation Telescope in New Mexico, USA, and the 2.5m Irenee du Pont Telescope in Las Campanas Observatory in Chile were used, covering the northern and southern skies respectively.
In the present work, we reanalyze the optical spectra for the elements Na, Mg, Al, Si, Ca, and Ti, and in the \textit{H}-band the elements
C, N, O, Mg, Al, Si, and Ca for the lines reported in Table \ref{linelist}.

\subsection{Previous calculations}

In the optical, we analyzed UVES spectra observed in the range 4800-6800 {\rm \AA}  based on
synthetic spectra calculations carried out with the code PFANT described in \citet{barbuy18b}\footnote{The code is available at http://trevisanj.github.io/PFANT.}. This code is an update of the original FANTOM or ABON2 Meudon code by M. Spite. Each model atmosphere was interpolated in the MARCS grids \citep{gustafsson08}. 

In the \textit{H}-band, the ASPCAP in its last release, DR17, issued stellar parameters obtained from a spectroscopic solution that
minimizes the errors in seven dimensions (T$_{\rm eff}$, log~{\tt g} , [Fe/H], v$_{\rm t}$,
[$\alpha$/Fe], [C/Fe], [N/Fe]), and yields the abundances of several elements. For DR17, the results were obtained with new spectral grids constructed using the Synspec \citep{hubeny17,hubeny21} spectral synthesis code that incorporates NLTE level populations for Na, Mg, K, and Ca \citep{osorio20}.

\subsection{Previous results}

In Table \ref{samplestars} we report the identification of the stars, their coordinates, and colours. Table \ref{parameters}  provides the stellar parameters derived in the  optical and using the APOGEE ASPCAP procedure
together with respective references.

Table \ref{abundanceapogee} lists the abundances derived for the sample stars in the APOGEE DR17 release. These data correspond to  the spectroscopic or 
so-called noncalibrated APOGEE abundances  adopted in this paper for comparison reasons.
These elemental abundances were obtained in the table (allStar-dr17-synspec$_{-}$rev1.fits)\footnote{
https://www.sdss4.org/dr17/irspec/spectro$_{-}$data/}  in the APOGEE
site.


Table \ref{abundanceuves} gives the abundances obtained with UVES for each sample star from \citet{barbuy13,barbuy14,barbuy15,barbuy16,barbuy18a,ernandes2018,ernandes20,friaca17,lecureur07,souza21} and \citet{swaelmen16}. 

\section{New calculations}\label{sec3}

We recomputed the abundances of C, N, O, Mg, Al, Si, and Ca in the \textit{H}-band,
and Na, Mg, Al, Si, Ca, and Ti  in the optical using the code
{\tt TURBOSPECTRUM} from \citet{alvarez98} and \citet{plez12}. 
Model atmosphere grids are from \citet{gustafsson08}. 
The solar abundances are from \citet{grevesse98}:  A(C)=8.55, A(N)=7.97,
from \citet{steffen15}: A(O)=8.76, and from \citet{smith21}: A(Na)=6.17, A(Mg)=7.53, 
A(Al)=6.37, A(Si)=7.51, A(Ca)=6.31, A(Ti)=4.90.

Table \ref{linelist} reports the lines in the \textit{H}-band that we verified in the spectra of the eight sample stars. The atomic line list employed is that from the APOGEE collaboration,
together with the molecular lines described in \citet{smith21}. The lines of the studied elements C, N, O, Na, Al, Mg, Si, and Ca (Ti is excluded) were recently examined by \citet{razera22} and \citet{barbuy23}.

In the optical, the lines measured are listed, for example, in \citet{barbuy16}. For the calculations, we adopted the line list from VALD3 \citep{ryabchikova15}.
The new VALD3 line list includes hyperfine structure when possible and needed.
The molecular lines include 
CH from \citet{masseron14}, 
CN from \citet{brooke14}, 
C2 from \citet{brooke13},
MgH from \citet{skory03}, and 
TiO, including the $\alpha$, $\beta$,  $\gamma$, $\gamma$', $\epsilon$,
$\delta$ (and $\phi$ in the NIR), from \citet{jorgensen94}.

\begin{table}
\small
\caption[4]{\label{linelist}
Lines studied in the \textit{H}-band. Oscillator strengths, log~gf, reported in \citet{smith13},  \citet{smith21}, or
given in VALD3 or Kurucz (1993)\tablefootnote{1995 Atomic Line Data (R.L. Kurucz and B. Bell) Kurucz CD-ROM No. 23. Cambridge, Mass.: Smithsonian Astrophysical Observatory.}, 
and NIST are reported.}
\begin{tabular}{l@{}cccccccccccccc}
\hline
\noalign{\smallskip}
\hbox{Species} & \hbox{$\lambda$} & \hbox{$\chi_{ex}$}  & \hbox{log~gf} & \hbox{log~gf} &\hbox{log~gf} &  \\
& \hbox{(\AA)} &\hbox{(eV)} & \hbox{Smith13} & \hbox{(VALD3)} & \hbox{(NIST)}   \\ 

\noalign{\smallskip}
\hline
\noalign{\smallskip}
\hbox{\ion{Mg}{I}} & 15740.716 & 5.931 & -0.262 & -0.323 & -0.212   \\
& 15748.988   & 5.932 & 0.276 & 0.049 & -0.338  \\
& 15765.842 & 5.933 & 0.504 & 0.320 & -0.337    \\
& 15879.521 & 5.946 & -1.248 & -2.102 & -1.998  \\
& 15886.261 & 5.946 & -1.555  &-1.742 & -1.521  \\
\hbox{\ion{Al}{I}}  & 16718.957 & 4.085 & 0.290 & 0.152 & 0.220  \\
& 16750.539 & 4.088 & --- & 0.408 & ---  \\
& 16763.359 & 4.087 & -0.524 & -0.550 & -0.480  \\
\hbox{\ion{Si}{I}}  & 15376.831 & 6.222 & -0.649 & -0.701 & ---  \\
& 15960.063 & 5.984 &  0.107 &  0.130 & 0.197   \\
& 16060.009 & 5.954 & -0.566 & -0.452 & -0.429   \\
& 16094.787 & 5.964 & -0.168 & -0.088 & -0.078   \\
& 16215.670 & 5.964 & -0.665 & -0.565 & -0.575  \\
& 16680.770 & 5.984 & -0.140 & -0.138 & -0.090   \\
& 16828.159 & 5.984 & -1.102 & -1.058 & -1.012  \\
\hbox{\ion{Ca}{I}} & 16150.763 & 4.532 & -0.229 & 0.362 & ---  \\
& 16155.236 & 4.532 & -0.619 & -0.758 & ---   \\
& 16157.365 & 4.554 &-0.208  & -0.219 & ---   \\
& 16197.075 & 4.535 & --- & 0.089 & ---   \\
& 16204.087 & 4.535 & --- & -0.627 & ---   \\
\noalign{\hrule\vskip 0.1cm}               
\end{tabular}
\end{table}

\begin{table*}
\centering
\caption{Sample stars: Identifications from internal name, Gaia, 2MASS and OGLE numbers. 
Star NGC6522-402361 is also identified as B-107 in \citet{barbuy14}, and stars OGLE-545277, - 82760 and -392918 are also identified as BW-b4, BW-b5 and BW-f6 in \citet{lecureur07}.}
\label{samplestars}
\begin{tabular}{l@{}cc@{}c@{}c@{}cccc}
\hline
\multicolumn{8}{c}{  HP1}                                       \\ \hline
ID internal   & Gaia  & 2MASS   & OGLE           & $\alpha_{2000}$ & $\delta_{2000}$ & V      & I      & H       \\ \hline
HP1-2 & 4058638335451540352 & 2M17310585-2958354 & ---  & 17:31:05.852    & \phantom{-}-29:58:35.5     & 16.982 & 14.332 & 11.268  \\ \hline
\\
\multicolumn{8}{c}{  NGC 6558}                                   \\ \hline
ID internal & Gaia     & 2MASS   & OGLE          & $\alpha_{2000}$ & $\delta_{2000}$ & V      & I      & H       \\ \hline
283   & 4048864158119516416 & 2M18102223-3145435  & ---  & 18:10:22.228    & \phantom{-}-31:45:43.340   & 15.883 & 14.378 & 12.564  \\
1160  & 4048864432989873408 & 2M18101768-3145246   & --- & 18:10:17.682    & \phantom{-}-31:45:24.570   & 15.586 & 14.063 & 12.000  \\ \hline
\\
\multicolumn{8}{c}{ NGC 6522}                                   \\ \hline
ID internal & Gaia     & 2MASS   & OGLE         & $\alpha_{2000}$ & $\delta_{2000}$ & V      & I      & H       \\ \hline
B-107 & 4050198110543819008    & 2M18033660-3002164 & \phantom{-}402361  & \phantom{-}18:03:36.59     & \phantom{-}-30:02:16.1     & 15.977 & 14.313 & 11.803  \\\hline
\\
\multicolumn{8}{c}{ Palomar 6}                                    \\ \hline
ID  internal & Gaia   & 2MASS    & OGLE          & $\alpha_{2000}$ & $\delta_{2000}$  & V      & I    & H 2MASS \\ \hline
401  & 4061102272286853888 & 2M17433806-2613426 & ---  & 17 43 38.05 & \phantom{-}-26 13 42.7 & 17.43  & 14.40   & 11.46   \\ 
\hline
\\
\multicolumn{8}{c}{ Bulge Field}                                            \\ \hline
ID internal & Gaia     & 2MASS    & OGLE        & $\alpha_{2000}$ & $\delta_{2000}$ & V      & I      & H 2MASS \\ \hline
BW-b4 & 4050186355212762624  & 2M18040535-3005529 & \phantom{-}545277  & \phantom{-}18 04 05.340    & 
\phantom{-}-30 05 52.50    & 16.95  & 14.42  & 11.803  \\
BW-b5 & 4050205119984310144  & 2M18041328-2958182 & 
\phantom{-}82760 & \phantom{-}18 04 13.270    & 
\phantom{-}-29 58 17.80    & 18.832 & 14.45  & 12.032  \\
BW-f6 & 4050196456967930240 & 2M18033691-3007047 & 
\phantom{-}392918 & \phantom{-}18 03 36.890    & 
\phantom{-}-30 07 04.30    & 18.387 & 14.272 & 12.043  \\
\hline
\end{tabular}%

\end{table*}


Below we give a brief description of the methods employed
in the analysis of optical and \textit{H}-band spectra 
in order to make clear that there are differences between them
due to the different lines available in each spectral region.

In the optical, initial parameters based on photometric indicators are adopted, and iteratively modified based on ionization and excitation equilibra of \ion{Fe}{I} and \ion{Fe}{II} lines of different ionization and excitation energies until the parameters converge. The excitation equilibrium requires a constant Fe abundance for \ion{Fe}{I} lines with different excitation energies $\chi_{\rm exc}$, and is affected mainly by iterating the value of the effective temperature. The ionization equilibrium requires similar Fe abundances from \ion{Fe}{I} and \ion{Fe}{II} lines, which are affected mainly by the gravity log~g value. Finally, the microturbulence velocity v$_{\rm t}$ is obtained by imposing a constant Fe abundance versus equivalent width (EW) of 
\ion{Fe}{I} lines. 

In the \textit{H}-band, the \ion{Fe}{I} lines have almost all of their excitation energies in the range 5.5 to 7.0 eV, which means that the concept of excitation equilibrium cannot be used as initial temperature indicator. Therefore,  initial indication for the effective temperature are photometric indices. The ASPCAP pipeline derives their noncalibrated stellar parameters, in this case from Data Release 17 (DR17),  using a spectroscopic solution that minimizes the 
chi-squared differences between the observations and the models in the following seven dimensions: T$_{\rm eff}$, log~g, [Fe/H], v$_{\rm t}$, [$\alpha$/Fe], [C/Fe], and [N/Fe]).
As we demonstrate below, the fit to molecular lines of CN, CH, CO, and OH
is a powerful tool for constraining stellar parameters, in particular
the effective temperature.

In the following sections, we carry out abundance derivations using the two sets of stellar parameters ---that is, those obtained with the APOGEE ASPCAP procedure and the detailed analyses with the optical spectra observed with UVES--- that are reported in Table \ref{parameters}.

\section{Stellar parameters}\label{sec4}

In this section, we compare effective temperature T$_{\rm eff}$, gravity log~g, metallicity [Fe/H], and microturbulence velocity v$_{\rm t}$ obtained from analyses of spectra in the optical and the \textit{H}-band, and element abundances derived in both cases.

\subsection{Comparison of stellar parameters from optical and near-infrared spectra}

The comparison of stellar parameters obtained from detailed analysis in the optical and the derivation adopted with the ASPCAP software in the \textit{H}-band can provide information as to the advantages and disadvantages of using each of these bands.

From Table \ref{parameters}, we make the following observations:

\begin{itemize}
\item For all stars, the microturbulence velocity is very low in the ASPCAP parameters.
\item Stars N6522-402361, OGLE-82760, and OGLE-392918 have very close parameters derived in the optical and \textit{H}-band.
\item Star HP1-2 has very similar  T$_{\rm eff}$ and log~g, but lower [Fe/H] and v$_{\rm t}$ from \textit{H}-band relative to optical.
\item Stars NGC6558-283, NGC6558-1160, Palomar6-401, and OGLE-545277 have  lower T$_{\rm eff}$ by about 300 K, higher log~g by about 0.7dex, lower v$_{\rm t}$ by about 0.8 km.s$^{-1}$, and relatively similar metallicities for the metal-poor stars and lower [Fe/H] by about 0.5dex for the metal-rich star OGLE-545277, that is, according to \textit{H}-band relative to optical analyses.
\end{itemize}

A discussion on which are the best stellar parameters is deferred to the following sections involving comparisons of element abundances, 
that in turn also  help us to discriminate the most likely results for each of the sample stars according to their stellar population group.


\begin{table*}[]
\centering
\caption{ ASPCAP (noncallibrated) spectroscopic parameters for the stars from APOGEE,
  and  from DGC23: \cite[hereafter DGD23]{gonzalez-diaz23} in the \textit{H}-band, and spectroscopic parameters from UVES from
B16: \cite{barbuy16}, B18: \cite{barbuy18a}, B14: \cite{barbuy14}, S21: \cite{souza21}, 
L07: \cite{lecureur07} in the optical. Parameters obtained from lower resolution FLAMES-GIRAFFE
spectra from B09: \cite{barbuy09} and Z08: \cite{zoccali08}, indicated by -G, are also reported.}\label{parameters}
\resizebox{\textwidth}{!}{%
\begin{tabular}{ccccccc|cccccc}
\hline
\multicolumn{7}{c|}{APOGEE}                                                                     & \multicolumn{6}{c}{UVES}                                                 \\ \hline
ID           & T$_{eff}$ (K) & log g & v$_{t}$ (km s$^{-1}$) & {[}Fe/H{]}         & Ref.  & SNR & T$_{eff}$ (K) & log g & v$_{t}$ (km s$^{-1}$) & {[}Fe/H{]} & Ref.  & SNR \\ \hline
HP1-2        & 4565          & 1.54  & 0.42                  & -1.229 $\pm$ 0.017 & DR17  & 122 & 4630          & 1.70  & 1.60       --           & -1.00      & B06   & 70  \\
N6558-283    & 4570          & 2.01  & 0.30                  & -1.043 $\pm$ 0.021 & DR17  & 46  & 4800          & 2.10  & 1.00    --              & -1.20      & B18   & 80  \\
N6658-1160   & 4454          & 1.94  & 0.20                  & -1.178 $\pm$ 0.019 & DR17  & 80  & 4900          & 2.60  & 1.30    --              & -1.15      & B18   & 80  \\
N6558-1160   & 4729          & 1.80  & 1.61                  & -1.15 $\pm$ 0.008  & DGD23 & --    &  --             &   --    &           --            &  --          &    --   & --    \\
N6522-402361 & 4993          & 2.43  & -0.11                 & -1.131 $\pm$ 0.022 & DR17  & 64  & 4990          & 2.00  & 1.40      --            & -1.12      & B14   & 110 \\
N6522-402361 &  --             &   --    &  --                     &        --            &   --    &   --  & 4990          & 2.10  & 1.40                  & -1.11      & B09-G &  --   \\
N6522-402361 &  --             &    --   &  --                     &           --         &    --   &  --   & 4950          & 2.00  & 1.40                  & -1.06      & Z08-G &  --   \\
Pal6-401a    & 4294          & 1.62  & 0.24                  & -0.911 $\pm$ 0.018 & DR17  & 61  & 4500          & 1.50  & 1.0                   & -1.00      & S21   & 50  \\
Pal6-401b    & 4396          & 1.80  & 0.24                  & -0.911 $\pm$ 0.018 & DR17  & 105 &      --        &     --  &              --         &     --       &  --     & --    \\
OGLE-545277  & 3955          & 2.23  & 0.06                  & 0.540 $\pm$ 0.010  & DR17  & 58  & 4300          & 1.40  & 1.4                   & 0.07       & L07   & 60  \\
OGLE-545277  &  --             &   --    &              --         &       --             &   --    &  --   & 4100          & 1.84  & 1.2                   & 0.35       & Z08-G & --    \\
OGLE-82760   & 3967          & 2.02  & 0.10                  & 0.317 $\pm$ 0.010   & DR17  & 87  & 4000          & 1.60  & 1.2                   & 0.17       & L07   & 60  \\
OGLE-82760   &  --             &     --  &             --          &     --               &  --    &   --  & 4300          & 1.87  & 1.5                   & 0.25       & Z08-G &  --   \\
OGLE-392918  & 4135          & 1.76  & 0.19                  & -0.260 $\pm$ 0.013 & DR17  & 69  & 4100          & 1.70  & 1.5                   & -0.21      & L07   & 60  \\
OGLE-392918  &  --             & --      &      --                 &    --                & --      &  --   & 4600          & 1.97  & 1.4                   & 0.05       & Z08-G &  --   \\ \hline
\end{tabular}%
}
\end{table*}

\begin{figure*}
    \centering
    \includegraphics[width=0.73\textwidth]{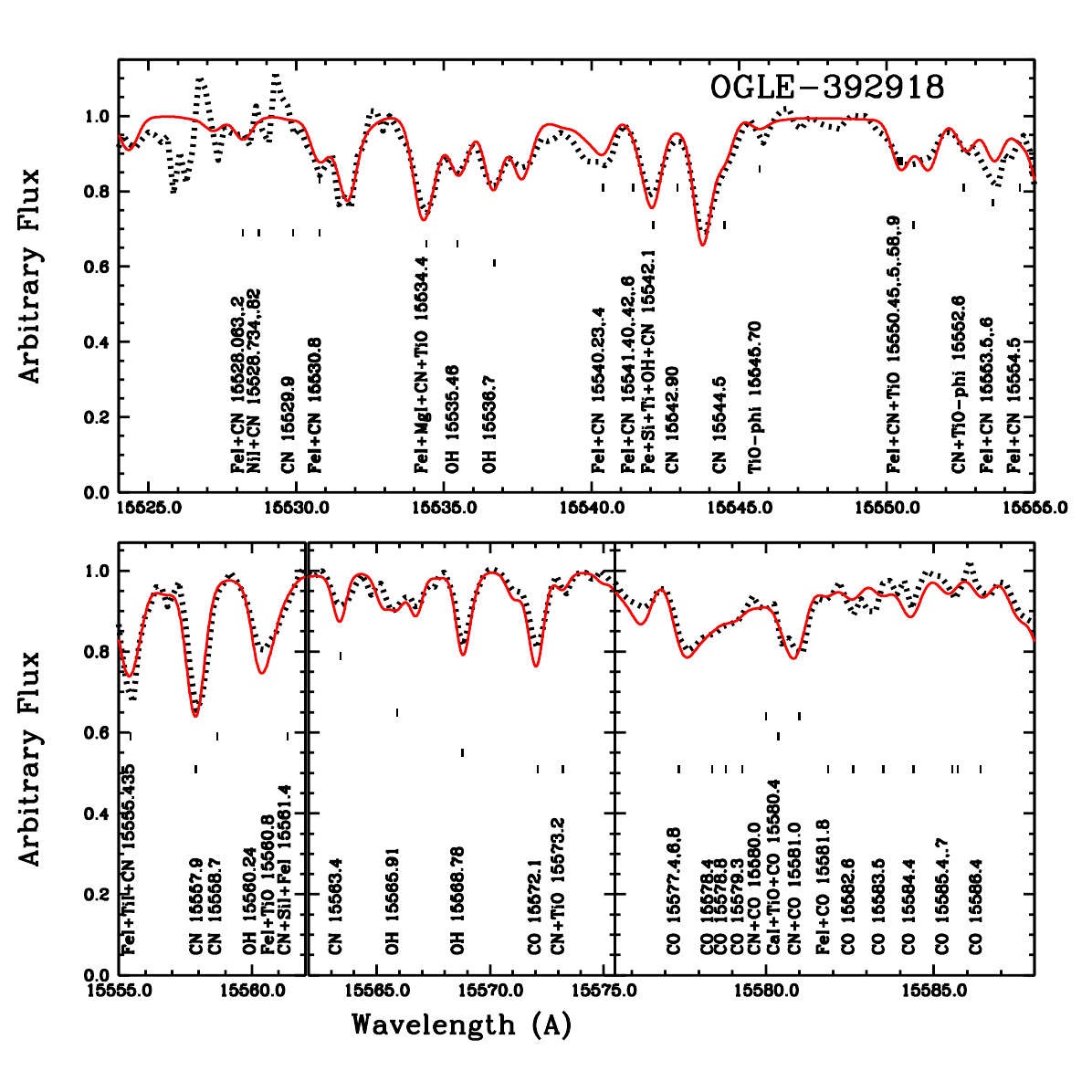}
    \caption{OGLE-393918: APOGEE spectrum fitted in the region 
    $\lambda\lambda$ 15525 - 15595 {\rm \AA} (black dotted lines) and synthetic spectra adopting
    the UVES stellar parameters (red lines). }
    \label{ogle39cnouves}
\end{figure*}

\begin{table*}
\centering
\caption{Abundance ratios [X/Fe] for each star from APOGEE ASPCAP DR17. 
For each element, the first line is the result adopted as calibrated by the APOGEE project, 
whereas the second line, in bold face, corresponds to the spectroscopic ASPCAP abundance. }
\label{abundanceapogee}
\begin{adjustbox}{width=\textwidth}
\begin{tabular}{ccccccccc}
\hline
Element & HP1-2 & 283   & 1160  & 402361 & 401   & 545277 & 82760  & 392918 \\ 
\hline
\multicolumn{9}{c}{APOGEE} \\
\hline
\multirow{2}{*}{C}  & $-$0.35 $\pm$ 0.04 & $-$0.16 $\pm$ 0.05 & $-$0.21 $\pm$ 0.04 & 0.45 $\pm$ 0.06 & $-$0.14 $\pm$ 0.03 & 0.074 $\pm$ 0.005 & 0.087 $\pm$ 0.006 & 0.158 $\pm$ 0.013  \\
& \textbf{$-$0.55 $\pm$ 0.04} & \textbf{$-$0.25 $\pm$ 0.05} & \textbf{$-$0.27 $\pm$ 0.05} & \textbf{0.17 $\pm$ 0.07} & \textbf{$-$0.27 $\pm$ 0.04} & \textbf{0.070 $\pm$ 0.011} & \textbf{0.082 $\pm$ 0.011} & \textbf{0.151 $\pm$ 0.019}  \\
\multirow{2}{*}{N}  & 0.89 $\pm$ 0.05 & 0.81 $\pm$ 0.06 & 0.07 $\pm$ 0.06 & 0.39 $\pm$ 0.08 & 0.34 $\pm$ 0.04 & 0.326 $\pm$ 0.010 & 0.282 $\pm$ 0.011  & 0.220 $\pm$ 0.021  \\
 & \textbf{1.02 $\pm$ 0.04} & \textbf{0.88 $\pm$ 0.05} & \textbf{0.16 $\pm$ 0.05} & \textbf{0.47 $\pm$ 0.06} & \textbf{0.47 $\pm$ 0.04} & \textbf{0.254 $\pm$ 0.013} & \textbf{0.258 $\pm$ 0.013} & \textbf{0.236 $\pm$ 0.022}  \\
\multirow{2}{*}{O}   & 0.26 $\pm$ 0.03 & 0.29 $\pm$ 0.05 & 0.34 $\pm$ 0.04 & 0.51 $\pm$ 0.07 & 0.32 $\pm$ 0.03 & 0.011 $\pm$ 0.009 & 0.038 $\pm$ 0.009 & 0.321 $\pm$ 0.016  \\
 & \textbf{0.19 $\pm$ 0.04} & \textbf{0.23 $\pm$ 0.06} & \textbf{0.33 $\pm$ 0.05} & \textbf{0.60 $\pm$ 0.07} & \textbf{0.32 $\pm$ 0.04} & \textbf{$-$0.015 $\pm$ 0.012} & \textbf{0.029 $\pm$ 0.012} & \textbf{0.277 $\pm$ 0.019}  \\
\multirow{2}{*}{Na}   & $-$0.23 $\pm$ 0.12  & $-$0.31 $\pm$ 0.14 & 0.22 $\pm$ 0.13  & $-$0.66 $\pm$ 0.16  & 0.16 $\pm$ 0.11  & 0.27 $\pm$ 0.03   & 0.15 $\pm$ 0.03   & 0.20 $\pm$ 0.06   \\
& \textbf{-1.24 $\pm$ 0.23} & \textbf{0.45 $\pm$ 0.29} & \textbf{0.05 $\pm$ 0.26} & \textbf{$-$0.87 $\pm$ 0.35} & \textbf{0.21 $\pm$ 0.21} & \textbf{0.33 $\pm$ 0.04} & \textbf{0.19 $\pm$ 0.04}  & \textbf{0.23 $\pm$ 0.09}  \\
\multirow{2}{*}{Mg}       & 0.21 $\pm$ 0.03 & 0.24 $\pm$ 0.03  & 0.29 $\pm$ 0.03  & 0.25 $\pm$ 0.03   & 0.28 $\pm$ 0.03  & $-$0.002 $\pm$ 0.013 & 0.059 $\pm$ 0.013  & 0.492 $\pm$ 0.018  \\
& \textbf{0.09 $\pm$ 0.03} & \textbf{0.20 $\pm$ 0.03} & \textbf{0.24 $\pm$ 0.03} & \textbf{$-$0.11 $\pm$ 0.03} & \textbf{0.23 $\pm$ 0.03} & \textbf{$-$0.035 $\pm$ 0.013}  & \textbf{$-$0.025 $\pm$ 0.013}  & \textbf{0.310 $\pm$ 0.019}  \\
\multirow{2}{*}{Al}      & 0.10 $\pm$ 0.03 & 0.28 $\pm$ 0.05  & $-$0.05 $\pm$ 0.04 & 0.26 $\pm$ 0.05   & 0.08 $\pm$ 0.04  & 0.03 $\pm$ 0.03   & 0.06 $\pm$ 0.03   & -0.14 $\pm$ 0.04 \\
& \textbf{0.27 $\pm$ 0.04} & \textbf{0.42 $\pm$ 0.05} & \textbf{0.13 $\pm$ 0.05} & \textbf{0.06 $\pm$ 0.05} & \textbf{0.29 $\pm$ 0.05} & \textbf{0.11 $\pm$ 0.04}  & \textbf{0.10 $\pm$ 0.03}  & \textbf{0.47 $\pm$ 0.04}  \\
\multirow{2}{*}{Si}       & 0.198 $\pm$ 0.022 & 0.15 $\pm$ 0.03  & 0.20 $\pm$ 0.03  & 0.39 $\pm$ 0.03  & 0.418 $\pm$ 0.024 & $-$0.059 $\pm$ 0.012 & -0.042 $\pm$ 0.012 & 0.279 $\pm$ 0.017 \\
& \textbf{0.23 $\pm$ 0.03} & \textbf{0.17 $\pm$ 0.04} & \textbf{0.24 $\pm$ 0.04} & \textbf{0.27 $\pm$ 0.04} & \textbf{0.42 $\pm$ 0.03} & \textbf{$-$0.017 $\pm$ 0.017}  & \textbf{0.011 $\pm$ 0.017}  & \textbf{0.228 $\pm$ 0.024}  \\
P       & 0.58 $\pm$ 0.15  & -0.29 $\pm$ 0.20 & -0.54 $\pm$ 0.17 & --     & 0.56 $\pm$ 0.14  & 0.07 $\pm$ 0.05  & 0.10 $\pm$ 0.05   & 0.29 $\pm$ 0.08   \\
\multirow{2}{*}{S}    & 0.51 $\pm$ 0.10  & 0.10 $\pm$ 0.13  & 0.24 $\pm$ 0.12  & 0.09 $\pm$ 0.13   & 0.75 $\pm$ 0.10  & $-$0.07 $\pm$ 0.03 & $-$0.05 $\pm$ 0.03  & 0.45 $\pm$ 0.06   \\
& \textbf{0.54 $\pm$ 0.13} & \textbf{0.37 $\pm$ 0.18} & \textbf{0.31 $\pm$ 0.16} & \textbf{0.09 $\pm$ 0.18} & \textbf{0.66 $\pm$ 0.15} & \textbf{$-$0.01 $\pm$ 0.04}  & \textbf{$-$0.03 $\pm$ 0.04}  & \textbf{0.35 $\pm$ 0.08}  \\
\multirow{2}{*}{K}      & 0.23 $\pm$ 0.09  & $-$0.05 $\pm$ 0.13 & 0.28 $\pm$ 0.11  & 0.28 $\pm$ 0.12   & 0.14 $\pm$ 0.10  & $-$0.15 $\pm$ 0.04  & 0.20 $\pm$ 0.04   & 0.07 $\pm$ 0.06   \\
& \textbf{0.14 $\pm$ 0.08} & \textbf{$-$0.03 $\pm$ 0.10} & \textbf{0.27 $\pm$ 0.09} & \textbf{0.38 $\pm$ 0.10} & \textbf{0.10 $\pm$ 0.09} & \textbf{$-$0.48 $\pm$ 0.04}  & \textbf{0.10 $\pm$ 0.04}  & \textbf{0.08 $\pm$ 0.06}  \\
\multirow{2}{*}{Ca}      & 0.14 $\pm$ 0.05  & 0.31 $\pm$ 0.06  & 0.28 $\pm$ 0.06  & $-$0.37 $\pm$ 0.07  & 0.22 $\pm$ 0.05  & $-$0.035 $\pm$ 0.010 & $-$0.010 $\pm$ 0.011 & 0.145 $\pm$ 0.022  \\
& \textbf{0.10 $\pm$ 0.04} & \textbf{0.34 $\pm$ 0.05} & \textbf{0.27 $\pm$ 0.05} & \textbf{$-$0.19 $\pm$ 0.06} & \textbf{0.28 $\pm$ 0.04} & \textbf{$-$0.046 $\pm$ 0.016}  & \textbf{$-$0.007 $\pm$ 0.016}  & \textbf{0.15 $\pm$ 0.03}  \\
\multirow{2}{*}{Ti}      & $-$0.01 $\pm$ 0.05 & 0.16 $\pm$ 0.07  & $-$0.01 $\pm$ 0.06 & 0.31 $\pm$ 0.09   & 0.01 $\pm$ 0.05  & --     & --     & --     \\
& \textbf{0.03 $\pm$ 0.05} & \textbf{0.22 $\pm$ 0.06} & \textbf{0.10 $\pm$ 0.05} & \textbf{0.29 $\pm$ 0.07} & \textbf{0.03 $\pm$ 0.05} & \textbf{0.730 $\pm$ 0.020}  & \textbf{0.421 $\pm$ 0.020}  & \textbf{0.09 $\pm$ 0.03}  \\
\multirow{2}{*}{TiII}    & 0.35 $\pm$ 0.11  & $-$0.03 $\pm$ 0.15 & $-$0.21 $\pm$ 0.13 & --     & 0.52 $\pm$ 0.13  & --     & --     & 0.34 $\pm$ 0.10  \\
& \textbf{0.22 $\pm$ 0.15} & \textbf{0.00 $\pm$ 0.20} & \textbf{0.24 $\pm$ 0.17} & \textbf{0.46 $\pm$ 0.22} & \textbf{0.30 $\pm$ 0.15} & \textbf{$-$0.05 $\pm$ 0.06}  & \textbf{0.04 $\pm$ 0.06}  & \textbf{0.15 $\pm$ 0.10}  \\
\multirow{2}{*}{V}      & 0.54 $\pm$ 0.13  & 0.27 $\pm$ 0.18  & $-$0.78 $\pm$ 0.15 & --     & $-$0.06 $\pm$ 0.13 & 0.08 $\pm$ 0.04  & 0.27 $\pm$ 0.04  & 0.02 $\pm$ 0.07  \\
& \textbf{0.33 $\pm$ 0.20} & \textbf{0.3 $\pm$ 0.3} & \textbf{$-$0.54 $\pm$ 0.23} & \textbf{$-$1.24 $\pm$ 0.32} & \textbf{$-$0.50 $\pm$ 0.19} & \textbf{$-$0.13 $\pm$ 0.05}  & \textbf{$-$0.14 $\pm$ 0.05}  & \textbf{$-$0.14 $\pm$ 0.10}  \\
\multirow{2}{*}{Cr}     & $-$0.17 $\pm$ 0.10 & 0.05 $\pm$ 0.14  & 0.12 $\pm$ 0.12  & $-$0.43 $\pm$ 0.15  & $-$0.72 $\pm$ 0.10 & $-$0.14 $\pm$ 0.03 & 0.09 $\pm$ 0.03   & 0.04 $\pm$ 0.06  \\
& \textbf{0.00$\pm$ 0.14} & \textbf{$-$0.03 $\pm$ 0.18} & \textbf{0.08 $\pm$ 0.16} & \textbf{$-$0.53 $\pm$ 0.21} & \textbf{$-$0.56 $\pm$ 0.13} & \textbf{0.03 $\pm$ 0.03}  & \textbf{0.06 $\pm$ 0.03}  & \textbf{$-$0.07 $\pm$ 0.06}  \\
\multirow{2}{*}{Mn}      & $-$0.28 $\pm$ 0.05 & $-$0.78 $\pm$ 0.07 & $-$0.30 $\pm$ 0.07 & $-$0.06 $\pm$ 0.08  & $-$0.12 $\pm$ 0.05 & --     & --     & $-$0.01 $\pm$ 0.03 \\
& \textbf{$-$0.45 $\pm$ 0.04} & \textbf{$-$0.77 $\pm$ 0.05} & \textbf{$-$0.41 $\pm$ 0.05} & \textbf{$-$0.28 $\pm$ 0.06} & \textbf{$-$0.24 $\pm$ 0.05} & \textbf{0.223 $\pm$ 0.016}  & \textbf{0.148 $\pm$ 0.016}  & \textbf{$-$0.09 $\pm$ 0.03}  \\
\multirow{2}{*}{Co}      & $-$0.83 $\pm$ 0.11 & $-$0.35 $\pm$ 0.15 & $-$0.17 $\pm$ 0.13 & $-$0.38 $\pm$ 0.18  & 0.12 $\pm$ 0.11  & 0.22 $\pm$ 0.03   & 0.16 $\pm$ 0.03  & 0.19 $\pm$ 0.05   \\
& \textbf{$-$0.94 $\pm$ 0.16} & \textbf{0.05 $\pm$ 0.26} & \textbf{$-$0.09 $\pm$ 0.20} & \textbf{0.9 $\pm$ 0.3} & \textbf{0.10 $\pm$ 0.17} & \textbf{0.27 $\pm$ 0.04}  & \textbf{0.14 $\pm$ 0.04}  & \textbf{0.15 $\pm$ 0.08}  \\
\multirow{2}{*}{Ni}      & 0.00 $\pm$ 0.04  & $-$0.10 $\pm$ 0.05 & 0.05 $\pm$ 0.05  & $-$0.26 $\pm$ 0.06  & 0.08 $\pm$ 0.04  & 0.013 $\pm$ 0.013  & 0.142 $\pm$ 0.014  & 0.111 $\pm$ 0.024  \\
& \textbf{$-$0.05 $\pm$ 0.03} & \textbf{$-$0.15 $\pm$ 0.05} & \textbf{0.04 $\pm$ 0.04} & \textbf{$-$0.20 $\pm$ 0.05} & \textbf{0.09 $\pm$ 0.04} & \textbf{0.011 $\pm$ 0.015}  & \textbf{0.096 $\pm$ 0.015}  & \textbf{0.070 $\pm$ 0.024}  \\
Cu      & 0.48 $\pm$ 0.11  & $-$1.13 $\pm$ 0.14 & 0.25 $\pm$ 0.13  & 0.77 $\pm$ 0.16  & 0.30 $\pm$ 0.10  & --     & --     & 0.21 $\pm$ 0.05 \\
\multirow{2}{*}{Ce}      & 0.12 $\pm$ 0.10  & 0.24 $\pm$ 0.13 & $-$0.25 $\pm$ 0.11 & 0.30 $\pm$ 0.15   & --    & --     & --     & --     \\
& \textbf{$-$0.10 $\pm$ 0.11} & \textbf{$-$0.38 $\pm$ 0.16} & \textbf{$-$0.58 $\pm$ 0.12} & \textbf{$-$0.07 $\pm$ 0.18} & \textbf{$-$0.27 $\pm$ 0.11} & \textbf{$-$0.20 $\pm$ 0.06}  & \textbf{$-$0.19 $\pm$ 0.06}  & \textbf{$-$0.33 $\pm$ 0.08}  \\
\hline
\end{tabular}
\end{adjustbox}
\end{table*}

\begin{table*}
\centering
\caption{Abundance ratios [X/Fe] obtained from UVES optical spectra for sample stars from \citet{barbuy13,barbuy14,barbuy15,barbuy16,barbuy18a}, \citet{ernandes2018,ernandes20,friaca17,lecureur07,souza21,swaelmen16}. Abundances in bold face were computed in the present work.}
\label{abundanceuves}
\begin{tabular}{rrrrrrrrr}
\hline
Element &\multicolumn{1}{c}{HP1-2 } & \multicolumn{1}{c}{283}  & \multicolumn{1}{c}{1160} & 
\multicolumn{1}{c}{402361} & \multicolumn{1}{c}{401}   & \multicolumn{1}{c}{545277} & \multicolumn{1}{c}{82760}
& \multicolumn{1}{c}{392918} \\ 
\hline
\multicolumn{9}{c}{UVES} \\
\hline
C    & 0.00$\pm$0.15  & 0.15$\pm$0.11 & 0.20$\pm$0.11 & 0.00$\pm$0.15  &$-$0.12$\pm$0.15  & $-$0.10$\pm$0.07 & 0.00$\pm$0.07  & 0.08$\pm$0.07   \\
N    & 0.50$\pm$0.15  & 0.80$\pm$0.11 & 1.00$\pm$0.11 & --             & 0.82$\pm$0.08  & 0.00$\pm$0.16  & 0.05$\pm$0.16  & 0.40$\pm$0.16   \\
O    & 0.30$\pm$0.15  & 0.40$\pm$0.11 & 0.50$\pm$0.11 & 0.50$\pm$0.12  & 0.45$\pm$0.15  & -0.10$\pm$0.08 & -0.10$\pm$0.08 & 0.20$\pm$0.08   \\
Na   & $-$0.02$\pm$0.05 & 0.15$\pm$0.11 & 0.00$\pm$0.11 & $-$0.22$\pm$0.05 & 0.10$\pm$0.15  & --- & --- & --- \\
Mg   & 0.15$\pm$0.06  & --            & 0.40$\pm$0.11 & 0.33$\pm$0.05  & 0.30$\pm$0.17  & --- & --- & --- \\
Al   & $-$0.28$\pm$0.12 & 0.30$\pm$0.12 & 0.00$\pm$0.12 & $-$0.30$\pm$0.12 & 0.40$\pm$0.17  & --- & --- & --- \\
Si   & 0.30$\pm$0.14  & --            & 0.20$\pm$0.10 & 0.17$\pm$0.10  & 0.41$\pm$0.17  & --- & --- & --- \\
Ca   & $-$0.04$\pm$0.15 & 0.00$\pm$0.13 & 0.16$\pm$0.13 & 0.16$\pm$0.05  & 0.34$\pm$0.18  &--- & --- & --- \\
TiI  & 0.07$\pm$0.13  & 0.15$\pm$0.15 & 0.15$\pm$0.15 & 0.03$\pm$0.10  & 0.32$\pm$0.18  &--- & --- & --- \\
TiII & 0.10$\pm$0.15  & 0.20$\pm$0.14 & 0.27$\pm$0.14 & 0.17$\pm$0.10   & ---           & --- & --- & --- \\
Sc   & $-$0.05$\pm$0.07 & $-$0.16$\pm$0.16 & $-$0.13$\pm$0.16 & $-$0.11$\pm$0.12&  ---  & ---   &  ---  &  ---   \\
V    & $-$0.22$\pm$0.09 & $-$0.01$\pm$0.03 & $-$0.01$\pm$0.03 & $-$0.06$\pm$0.06&  ---  & ---   &  ---  &  ---   \\
Mn   & $-$0.57$\pm$0.05 & $-$0.45$\pm$0.05 & $-$0.45$\pm$0.05 & $-$0.55$\pm$0.05&  ---  & 0.00$\pm$0.05  & 0.00$\pm$0.05  & 0.00$\pm$0.05   \\
Co   & ---            & ---            & ---            &  ---          &  ---  & -0.13$\pm$0.09 & -0.02$\pm$0.09 & 0.00$\pm$0.09   \\
Cu   &  $-$1.0$\pm$0.15 & $-$0.50$\pm$0.16 & $-$0.70$\pm$0.16 &$-$0.45$\pm$0.12 &  ---  & $-$0.30$\pm$0.12 & $-$0.35$\pm$0.12 & $-$0.50$\pm$0.12  \\ 
Zn   & 0.30$\pm$0.15  & $-$0.10$\pm$0.15 & 0.20$\pm$0.15  & 0.1$\pm$0.05  &  ---  &  0.00$\pm$0.05 & $-$0.30$\pm$0.05 &  0.15$\pm$0.05   \\
Y    & $-$0.15$\pm$0.15 & 0.75$\pm$0.19  & 0.56$\pm$0.19  & 0.32$\pm$0.25 & 0.09$\pm$0.15  &  ---  &  ---  &  ---   \\
Zr   & ---            & ---            & ---            & 0.20$\pm$0.20 & 0.41$\pm$0.35  &  ---  &  ---  &  ---   \\
Ba   & 0.65$\pm$0.16  & 0.00$\pm$0.10  & 0.15$\pm$0.10  & 0.45$\pm$0.23 & 0.49$\pm$0.13  & 0.16$\pm$0.21 & 0.36$\pm$0.21  & 0.06$\pm$0.21  \\
La   & $-$0.15$\pm$0.12 & ---            & ---            & 0.20$\pm$0.16 & 0.24$\pm$0.16  & ---   & 0.16$\pm$0.11  & 0.16$\pm$0.11   \\
Ce   & ---    & ---   & ---            & ---            & ---           & 0.04$\pm$0.11  & $-$0.34$\pm$0.11 & 0.08$\pm$0.11   \\
Nd   & ---    & ---   & ---            & ---            & ---           & $-$0.46$\pm$0.08 & $-$0.22$\pm$0.08 & $-$0.08$\pm$0.08   \\
Eu   & 0.40$\pm$0.15  & 0.80$\pm$0.12  & 0.30$\pm$0.12  & 0.40$\pm$0.12 & 0.58$\pm$0.11  & 0.26$\pm$0.10  & $-$0.20$\pm$0.10 & 0.26$\pm$0.10  \\
\hline
\end{tabular}
\end{table*}

\subsection{CNO abundances from Turbospectrum}

We first fitted the full region $\lambda\lambda$ 15525 - 15595 {\rm \AA,} which contains molecular lines of CO, OH, and CN, 
 as described in \citet{barbuy21b} and \citet{razera22}.
 The most important and clean of these features is the CO bandhead at  $\lambda\lambda$ 15578 - 15579 {\rm \AA}, which is sensitive to the C abundance, recalling that the less abundant species is the one that controls the feature. 
 After fitting this CO feature, we checked and fitted the OH lines at 15535.46, 15536.7, 15565.91, and 15568.78 {\rm \AA}  by changing the O abundance. Finally, we inspected the CN features along this region of the spectrum, which provides preliminary information on the N abundance. 
Figure \ref{ogle39cnouves} shows the fit to this region for star OGLE-3932918 computed with the stellar parameters from the
optical analysis.
For verification and fine tuning, a few OH, CN, and CO lines in other regions are fitted, as shown in \citet{razera22}. Table \ref{turbocno} shows the CNO abundances from {\tt TURBOSPECTRUM} calculations of APOGEE spectra 
for both sets (optical and \textit{H}-band) of stellar parameters.

For NGC6558-283, NGC6558-1160, and Pal6-401, it appears that the effective temperatures deduced in the optical 
are too high \citep{barbuy18c}, and are incompatible with the strength of the molecular lines
of CO and OH in particular. For NGC6558-1160, the ASPCAP effective temperature (T$_{\rm eff}$ = 4454 K)
appears suitable, and those somewhat higher ones (T$_{\rm eff}$ = 4729 K) by 
  \citet{gonzalez-diaz23} (DGD23) are acceptable.
 We note that DGD23 derived stellar parameters from VVV
photometry, and these are closer to the ASPCAP-calibrated parameters (T$_{\rm eff}$ = 4628.5 K, log~g = 1.813, 
v$_{\rm t}$ = 0.2145 km.s$^{-1}$).

We conclude that the stellar parameters obtained from ASPCAP, which correspond to  a spectroscopic solution 
that minimizes the chi-squared differences between the observations and the models in seven dimensions (T$_{\rm eff}$, log~g, [Fe/H], v$_{t}$,
[$\alpha$/Fe], [C/Fe],  and [N/Fe]), lead to reliable parameters that are constrained by the strength of molecular lines.
In the optical, \ion{Fe}{I} of different excitation energies are
used to constrain the effective temperature, together with
\ion{Fe}{I}  and \ion{Fe}{II} lines used to impose ionization
equilibrium, and the goodness of the procedure depends on reliable equivalent
widths of \ion{Fe}{II,} which are often difficult to obtain, in particular given the blends and noise
in the spectra of metal-poor stars.

The ASPCAP stellar parameters and CNO abundances are clearly more suitable in the
case of NGC6568-1160, for which the APOGEE spectra have a good S/N,
although this is not fully applicable to NGC6558-283 and NGC6522-402361, which have rather low S/N APOGEE spectra.
We consider this to be an important point concerning the ASPCAP derivation
of stellar parameters, as well as the C, N, and O abundances. In other words,
the fit to molecular lines appears to be a very suitable technique for deriving
reliable effective temperatures.
For the optical, we assumed the same C, N, and O abundances derived from the \textit{H}-band, which is more reliable for deriving C, N, and O given
that, in the optical, the C and N indicators ---that is, the C$_2$ and CN bandheads--- are very weak.

\begin{table*}
\centering
\caption{C, N, and O abundances from {\tt TURBOSPECTRUM} calculations of APOGEE spectra for both sets (optical and \textit{H}-band) of stellar parameters.}
\label{turbocno}
\begin{tabular}{c |  @{}c @{}c |  @{}c @{}c  |  @{}c  @{}c @{}c   |   @{}c@{}c | @{}c@{}c | @{}c@{}c |   @{}c@{}c |  @{}c@{}c |} \\ 
\hline
\multicolumn{18}{c}{Turbospectrum \textit{H}-band} \\
Model & \phantom{-}APO & \phantom{-}UVES & \phantom{-}APO & \phantom{-}UVES & \phantom{-}APO & \phantom{-}UVES & \phantom{-}DGD & \phantom{-}APO & \phantom{-}UVES & \phantom{-}APO & \phantom{-}UVES& \phantom{-}APO & \phantom{-}UVES& \phantom{-}APO & \phantom{-}UVES& \phantom{-}APO & \phantom{-}UVES \\
Star &\multicolumn{2}{c}{HP1-2 } & \multicolumn{2}{c}{283}  & \multicolumn{3}{c}{1160} & 
\multicolumn{2}{c}{402361} & \multicolumn{2}{c}{401}   & \multicolumn{2}{c}{545277} & \multicolumn{2}{c}{82760}
& \multicolumn{2}{c}{392918} \\ \hline
C   & $-$0.15 & $-$0.15   &0.12 &0.35 & 0.00&0.00  &0.15 &0.08 & 0.08: & 0.00 &0.45 &0.00 &+0.10  & $-$0.30 & $-$0.10  & $-$0.15 & -0.20   \\
N   & 0.60 & 0.40    &0.35 &0.80 & 0.00&0.10 &0.10& 0.30 & 0.30:   & 0.20 &0.2 &0.00 &+0.40 & 0.30 & +0.05  & 0.15 & 0.10   \\
O   & 0.27 & 0.45    &0.45 &0.50:& 0.30 &0.30 &0.40 & 0.42 & 0.50:  & 0.25&0.50: &0.00 &+0.50 & 0.30 & -0.05  & 0.20 &  0.10  \\
\hline
\end{tabular}
\end{table*}


\subsection{The atomic lines of the odd-Z Na, Al, and $\alpha$-elements Mg, Ca, Si, and Ti}

The sample stars are in the metallicity range $-$1.1 $\simless$ [Fe/H] $\simless$ +0.3.
Below, we describe which region, that is, the \textit{H}-band or optical, provides better indicators for each of the elements studied here.

{\it Magnesium:} \ion{Mg}{I}  lines are well-fit for all stars for the given abundances both from ASPCAP and UVES analyses.
For more metal-poor stars, most of these lines will become faint, and in those cases the stronger lines,
such as \ion{Mg}{I} 5528 {\rm \AA} and the  \ion{Mg}{I} triplet at 5167-5183 {\rm \AA,} might be appropriate.

{\it Silicon:} 

\begin{figure*}
\centering
\floatbox[{\capbeside\thisfloatsetup{capbesideposition={right,center},capbesidewidth=4.3cm}}]{figure}[\FBwidth]
{\caption{\textit{H}-band \ion{Si}{I} lines shown for the star NGC6522-402361.
    The APOGEE observed spectrum  (black line) is compared with synthetic spectra computed adopting the
    UVES stellar parameters and [Si/Fe]=+0.30 (green lines). }
    \label{n652240siuves}}
{\includegraphics[width=5.3in]{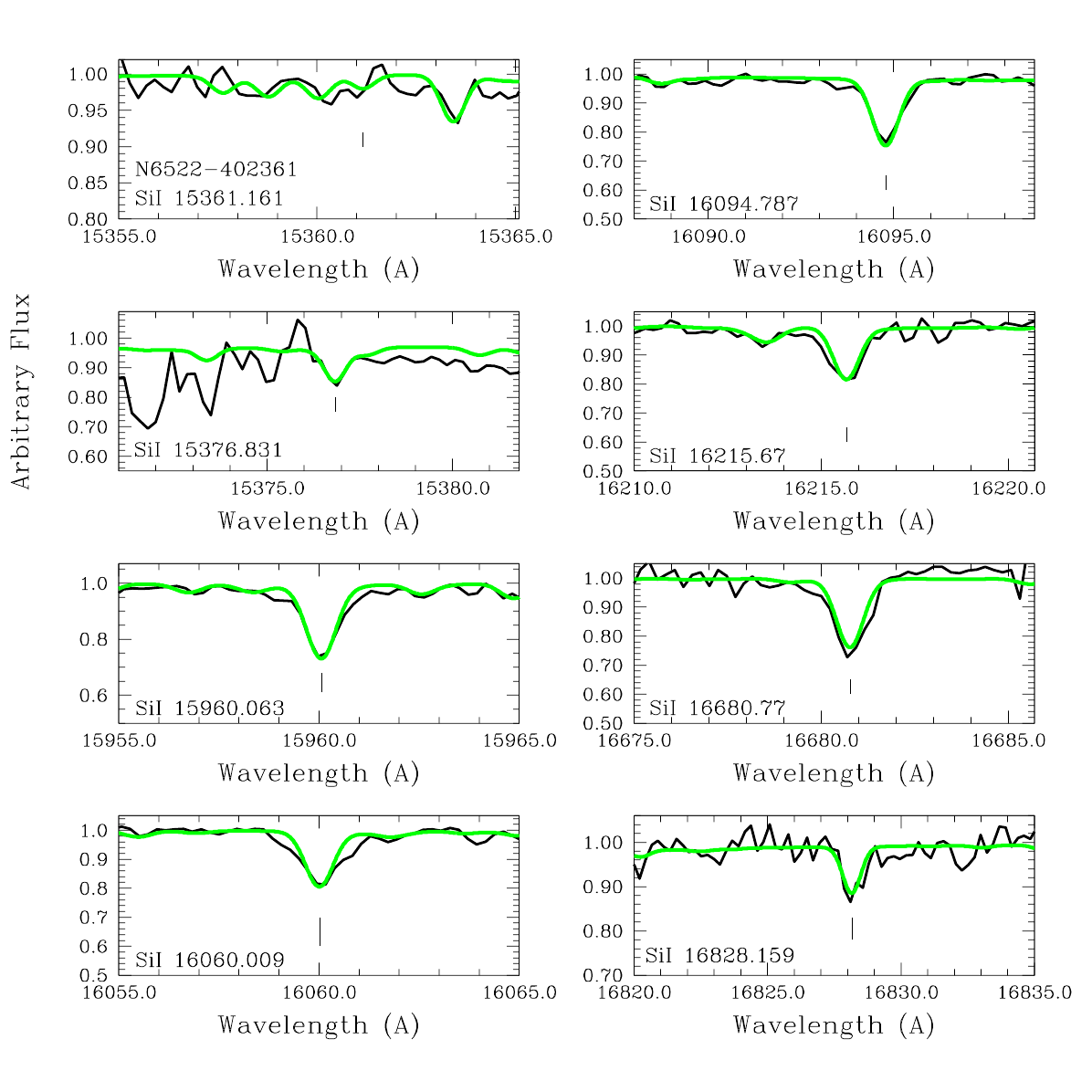}}
\end{figure*}

Clearly the \ion{Si}{I} lines in the \textit{H}-band are very suitable as Si abundance indicators; these comprise seven very good lines (we discard the \ion{Si}{I} 15361.161 {\rm \AA} line, which is weak).
In the optical, the lines are fainter for the metal-poor stars, but still good indicators,
and very suitable for the metal-rich stars. The \textit{H}-band \ion{Si}{I} lines, which come from
\citet{smith13,smith21}, are shown in Figure \ref{n652240siuves} for the star NGC6522-402361. In this figure, the line  observed \ion{Si}{I} 16060.009 {\rm \AA,} appears broader than the computed synthetic spectrum because of the unidentified lines on both sides of the \ion{Si}{I} line.

\begin{figure*}
\centering
\floatbox[{\capbeside\thisfloatsetup{capbesideposition={right,center},capbesidewidth=4.4cm}}]{figure}[\FBwidth]
{\caption{Optical \ion{Ca}{I} lines shown for the star NGC6558-1160. The UVES observed spectrum  (black line) is compared with synthetic spectra computed adopting the
    UVES and APOGEE stellar parameters,  with [Ca/Fe]=0.15 and 0.25 for UVES and APOGEE stellar parameters,
    respectively (blue and green lines are coincident). }
    \label{n65581160caturbo}}
{\includegraphics[width=5.15in]{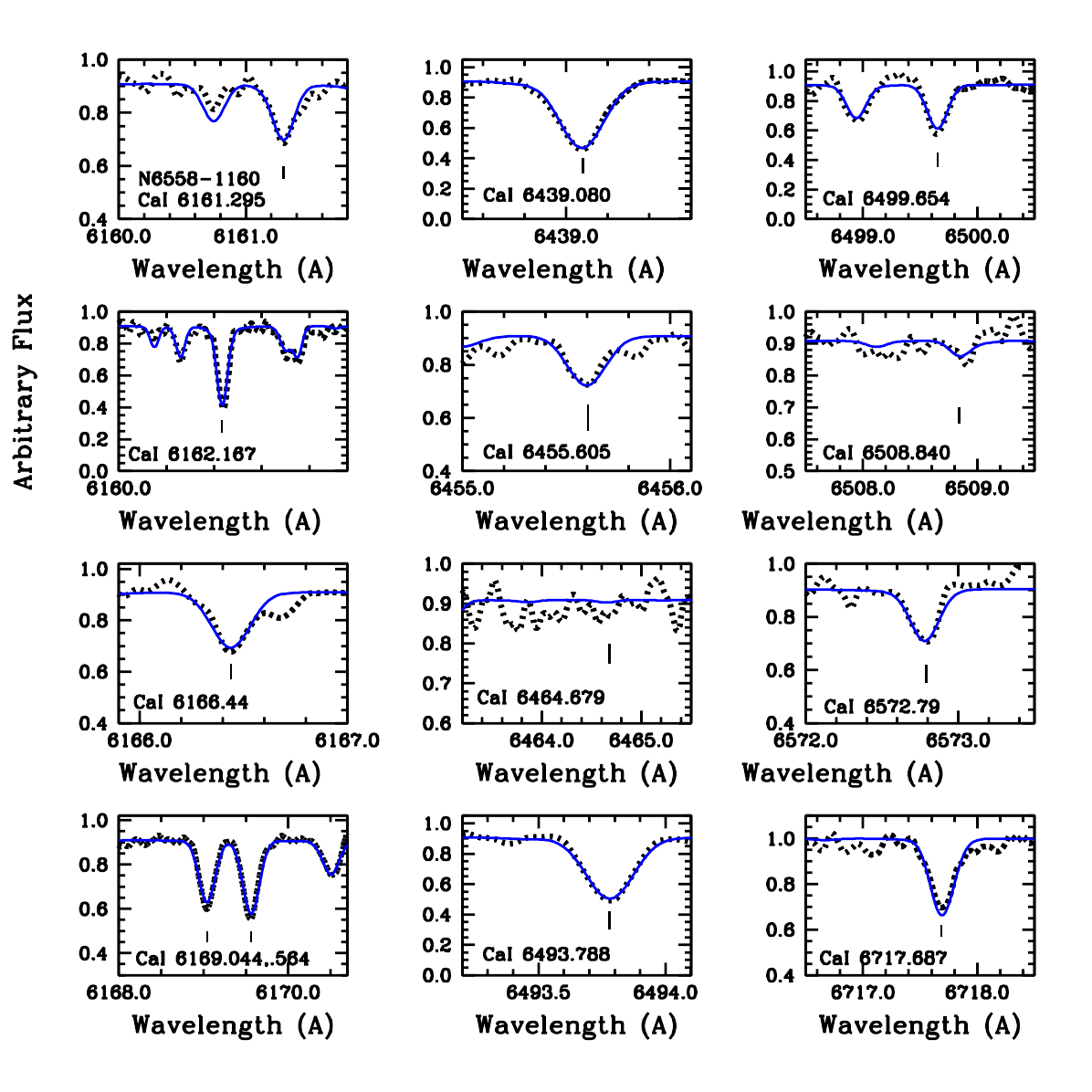}}
\end{figure*}

{\it Calcium:} Among the five \ion{Ca}{I} lines considered in the \textit{H}-band, only \ion{Ca}{I} 16155.235 {\rm \AA} is sufficiently sensitive to Ca abundance. Therefore, we conclude that the Ca abundances from the optical are more reliable than those from the \textit{H}-band.
Figure \ref{n65581160caturbo} shows the \ion{Ca}{I} lines in the optical for star NGC6558-1160.
In this figure, the two very weak lines, \ion{Ca}{I} 6464.679 and 6508.840 {\rm \AA,} are not useful in this metal-poor
star but they are useful in metal-rich stars.

{\it The odd-Z elements Na and Al:} As discussed in \citet{barbuy23}, the Na lines in the \textit{H}-band are not  reliable. As the lines in the optical are reliable, this band should be favoured for this element.
\ion{Al}{I} 16718.957, 16750.539 and 16763.358 {\rm \AA} are very suitable lines, whereas, in the optical, the \ion{Al}{I} 6696 and 6698 {\rm \AA} are strong in the field  metal-rich stars, but
rather weak in the metal-poor stars. Table \ref{turbonir} reports the abundances derived from atomic lines
of  \ion{Mg}{I}, \ion{Al}{I}, \ion{Si}{I}, and \ion{Ca}{I} in the \textit{H}-band
 computed with {\tt TURBOSPECTRUM} based on the lines
reported in Table \ref{linelist}.
Table \ref{turbovis} reports the abundances derived from atomic lines
of \ion{Na}{I}, \ion{Mg}{I}, \ion{Al}{I}, \ion{Si}{I}, \ion{Ca}{I}, \ion{Ti}{I,} and \ion{Ti}{II} in the optical, which are
 computed with {\tt TURBOSPECTRUM} based on the lines
reported in, for example, \citet{barbuy16}.

\begin{table*}
\centering
\caption{{\tt TURBOSPECTRUM} calculations of APOGEE spectra based on both APOGEE and UVES stellar parameters.}
\label{turbonir}
\begin{tabular}{c  |  @{}c @{}c |  @{}c @{}c  |  @{}c @{}c   |   @{}c@{}c | @{}c@{}c | @{}c@{}c |   @{}c@{}c | 
@{}c@{}c |} \\ 
\hline
\multicolumn{17}{c}{Turbospectrum} \\
Model & \phantom{-}APO & \phantom{-}UVES & \phantom{-}APO & \phantom{-}UVES & \phantom{-}APO & \phantom{-}UVES & 
\phantom{-}APO & \phantom{-}UVES & \phantom{-}APO & \phantom{-}UVES& \phantom{-}APO & \phantom{-}UVES& \phantom{-}APO & \phantom{-}UVES& 
\phantom{-}APO & \phantom{-}UVES \\
Star &\multicolumn{2}{c}{HP1-2 } & \multicolumn{2}{c}{283}  & \multicolumn{2}{c}{1160} & 
\multicolumn{2}{c}{402361} & \multicolumn{2}{c}{401}   & \multicolumn{2}{c}{545277} & \multicolumn{2}{c}{82760}
& \multicolumn{2}{c}{392918} \\ \hline

Mg & 0.30 & 0.20 & 0.20 &  0.32   & 0.25&  0.40  &  & 0.30 
&0.28&  0.20  & 0.00 &  0.00 &   0.00 &  -0.15 &  0.10 &  0.10   \\
Al  &  0.40  & 0.23 & 0.75 & 0.90  & 0.00 &  0.40 & --- & ---
&0.08&  0.35 & 0.30 &   0.65 & 0.70 &  0.50  &  0.70 &   0.35   \\
Si  &  0.50  & 0.20  &  0.40   & 0.30 &  0.20  & 0.40 &  0.30 &  0.30 &
0.42 & 0.45 &  0.00  & 0.06 &  0.15 &   0.05 &  0.35 &   0.15 \\
Ca  &  0.20  &  -0.05  &  ---  &   --- &  0.20 &  0.50  &  0.20 &    0.25 &
0.22&  0.30  &  0.40 &  0.35  &  0.50  &  0.05  &  0.40 &   0.15       \\
Ti  & ---  & ---  & ---  &  --- &---& ---   & ... &  ---  
&---&   ---  & --- &  ---    & --- &  --- & --- &  --- \\
\hline
\end{tabular}
\end{table*}

\begin{table*}
\centering
\caption{{\tt TURBOSPECTRUM }calculations of  lines in the UVES optical spectra based on both APOGEE and UVES stellar parameters.
}
\label{turbovis}
\begin{tabular}{c  |  @{}c @{}c |  @{}c @{}c  |  @{}c @{}c   |   @{}c@{}c | @{}c@{}c | @{}c@{}c |   @{}c@{}c | 
@{}c@{}c |} \\ 
\hline
\multicolumn{17}{c}{Turbospectrum} \\
Model & \phantom{-}APO & \phantom{-}UVES & \phantom{-}APO & \phantom{-}UVES & \phantom{-}APO & \phantom{-}UVES & \phantom{-}APO& \phantom{-}UVES & 
\phantom{-}APO & \phantom{-}UVES& \phantom{-}APO & \phantom{-}UVES& \phantom{-}APO& \phantom{-}UVES& \phantom{-}APO& \phantom{-}UVES \\
Star &\multicolumn{2}{c}{HP1-2 } & \multicolumn{2}{c}{283}  & \multicolumn{2}{c}{1160} & 
\multicolumn{2}{c}{402361} & \multicolumn{2}{c}{401}   & \multicolumn{2}{c}{545277} & \multicolumn{2}{c}{82760}
& \multicolumn{2}{c}{392918} \\ \hline
Na   & 0.50 &  0.20 &   0.50 & 0.30&  0.25 & 0.45&  0.05 & 0.10 
& -0.20 &  0.00 &  0.90  & 0.80 &  0.7 &  0.40  &   0.60&  0.35   \\
Mg  &  0.30 &0.20 & 0.20&  0.20   & 0.25&  0.40 &  0.30 &  0.30 
& 0.23&  0.21  & --- &  0.00  &  0.00 &  -0.15  &  0.20: &  0.10:    \\
Al &  0.45  & 0.30 &  0.60 &  0.35  & ---&   --- &  0.30 & 0.35
& 0.15&  0.35 & 0.80 &  0.65  &  0.50   &  0.50   &  0.80 &  0.35  \\
Si  &  0.40& 0.20&  0.45   & 0.35&  0.40  & 0.40&  0.10 & -0.10 & 
 0.40  & 0.45&  0.00  & 0.00&  0.20 &  0.05 &  0.35 & 0.15 \\
Ca  &  0.25 & -0.05  & 0.15 &  0.25 & 0.25& 0.35 &  -0.15 &   0.00 &
 0.0: &  0.0:   & 0.55   & 0.40  &  0.40 & 0.05  &  $>$0.8  &  0.15       \\
Ti &  0.30 & 0.15 & 0.20 & 0.45 & 0.15 &  $>$0.40 &  0.25 &  0.25  & 0.10 &   0.20&
 $>$0.5 &  0.40   &  $>$0.50 &  0.30 &  $\geq$0.70 &   0.0 \\
\hline
\end{tabular}
\end{table*}

\begin{table*}
\centering
\caption{Abundances derived from APOGEE and UVES spectra with {\tt TURBOSPECTRUM},
adopting the best-suited stellar parameters, as explained in Sect. 4.1.}
\label{niropt}
\begin{adjustbox}{width=\textwidth}
\begin{tabular}{crr | rr | rr |rr | rr | rr | rr | rr}
\hline
Model & \textit{H}-band & Opt & \textit{H}-band & Opt & \textit{H}-band & Opt & \textit{H}-band & Opt & \textit{H}-band & Opt & \textit{H}-band & Opt & \textit{H}-band & Opt & \textit{H}-band & Opt \\
Star &\multicolumn{2}{c}{HP1-2 } & \multicolumn{2}{c}{283}  & \multicolumn{2}{c}{1160} & 
\multicolumn{2}{c}{402361} & \multicolumn{2}{c}{401}   & \multicolumn{2}{c}{545277} & \multicolumn{2}{c}{82760}
& \multicolumn{2}{c}{392918} \\ \hline
C    &$-$0.15&  ---   &0.12&  --- & 0.00&  --- &0.08& ---   & 0.00 & --- & 0.10& ---  & $-$0.10&  ---  & $-$0.20&  ---   \\
N    &0.40 & ---  &0.35&  ---    & 0.00  &  --- & 0.30 &  --- & 0.20 & --- & 0.40  & ---   & 0.05 & ---  & 0.10& ---   \\
O    &0.45 & --- &0.45&  --- & 0.30& --- &0.50& ---   &0.25 & ---  & 0.50& --- &-0.05&  --- & +0.10 &  ---   \\
Na   &---& $-$0.20 &---& 0.50 &  --- &0.25& --- & 0.08 &  0.03  &$-$0.10 & 0.70 &0.80& 0.37  &0.40 & 0.00 & 0.35  \\
Mg   &0.21& 0.15   &0.24& -- & 0.25 & 0.25 & 0.30 & 0.30  & 0.24 & 0.23  & 0.00& 0.00 & $-$0.15 & $-$0.15  & 0.10 &  0.10   \\
Al   & 0.32 & 0.37 & 0.90 & 0.60  & 0.40 & --- & 0.08 & 0.32  & 0.32 &  0.25 & 0.70 & 0.65 & 0.70 & 0.50 & 0.35 & 0.35  \\
Si   &0.35& 0.30   &0.40 & 0.45 & 0.20&  0.40 & 0.30 & 0.00 & 0.30 & 0.46& 0.06  & 0.00 & 0.05  &0.05 & 0.15 & 0.15 \\
Ca   &0.12& 0.10  &---&  0.15 & 0.20& 0.25 & 0.22 & $-$0.07 & 0.26& 0.00  & 0.35 & 0.48 & 0.05 &0.05      &0.15& 0.15       \\
Ti  &---& 0.22 &--- & 0.20 & --- & 0.15& --- & 0.25  & ---& 0.15  & ---& 0.40 & ---&  0.30 &---& 0.00 \\
\hline
\end{tabular}
\end{adjustbox}
\end{table*}

\section{Discussion}\label{sec5}

In this work, we fitted the lines in the optical and \textit{H}-band and compared the results between these two sets. If the resulting abundances of some lines needed for the fits are too overabundant or too deficient, this indicates that the stellar parameters are not suitable.


\subsection{Stellar parameters}

Adopting the stellar parameters from the optical, the stars NGC6558-283 and NGC6558-1160 show weaker OH lines than observed, even considering a high oxygen abundance of [O/Fe]=0.5 - 0.6 (Table 
\ref{turbocno}), indicating that the effective temperature is probably too hot, and therefore that the ASPCAP parameters, 
which give considerably lower effective temperatures, are more suitable. The same applies to star Pal6-401, but with less discrepant effective temperatures between ASPCAP and optical.

The excessively high temperature derived from the optical 
is probably due to difficulties in measuring Fe lines because of blends, continuum placement,
 and noise. In conclusion, for the two
stars in NGC~6558, the APOGEE stellar parameters, as well as the parameters from \citet{gonzalez-diaz23}, are more suitable. 

For the star OGLE-545277, the ASPCAP parameters show a very high metallicity of [Fe/H]=+0.54, which is probably an overestimate, and this leads to a low oxygen abundance of [O/Fe]=$-$0.35; the optical parameters give a reasonable metallicity of [Fe/H]=+0.07. 

For OGLE-82760, despite very similar parameters, the low microturbulence velocity from ASPCAP requires a somewhat lower O abundance.
The abundances required to fit the lines Mg, Si, Ca, and Ti are far too strong ([X/Fe/]$>$0.40) for the fits of three OGLE stars (OGLE-545277, OGLE-82760, and OGLE-392918) using the ASPCAP stellar parameters, which is not expected for solar-metallicity stars. Therefore, the optical parameters given in \citet{zoccali06} and \citet{lecureur07} are clearly more suitable than the ASPCAP parameters for these three
solar-metallicity stars.
The stars HP1-2, NGC 6522-402361, and, to a lesser extent, Palomar6-401 have similar parameters derived
from the optical and \textit{H}-band, and they appear suitable.

\subsection{Discussion on atomic lines in the \textit{H}-band and optical}

\ion{Mg}{I}, \ion{Al}{I}, and \ion{Si}{I} lines are suitable in the \textit{H}-band and optical. The listed lines may be limited for more metal-poor stars in the case of Mg and Si, and of Al in the optical. Si has better (stronger) lines in the \textit{H}-band, whereas the lines are rather weak in the
optical.

\ion{Ca}{I} lines considered in the \textit{H}-band are not fully suitable because of blends. The lines in the optical are more suitable.

As discussed in \citet{razera22} and \citet{barbuy23}, \ion{Na}{I}, and \ion{Ti}{I} in the \textit{H}-band, are too blended ---or too weak in the case of Na. For these two elements, the optical is better suited.

Following these conclusions, and adopting the more suitable stellar parameters, the final abundances are given in Table \ref{niropt}.
This table shows the remarkable agreement between abundances, within errors, derived from \textit{H}-band and optical, once the correct stellar parameters are chosen.

\subsection{Comparison with bulge data}

\begin{figure}
    \centering
    \includegraphics[width=8.5cm]{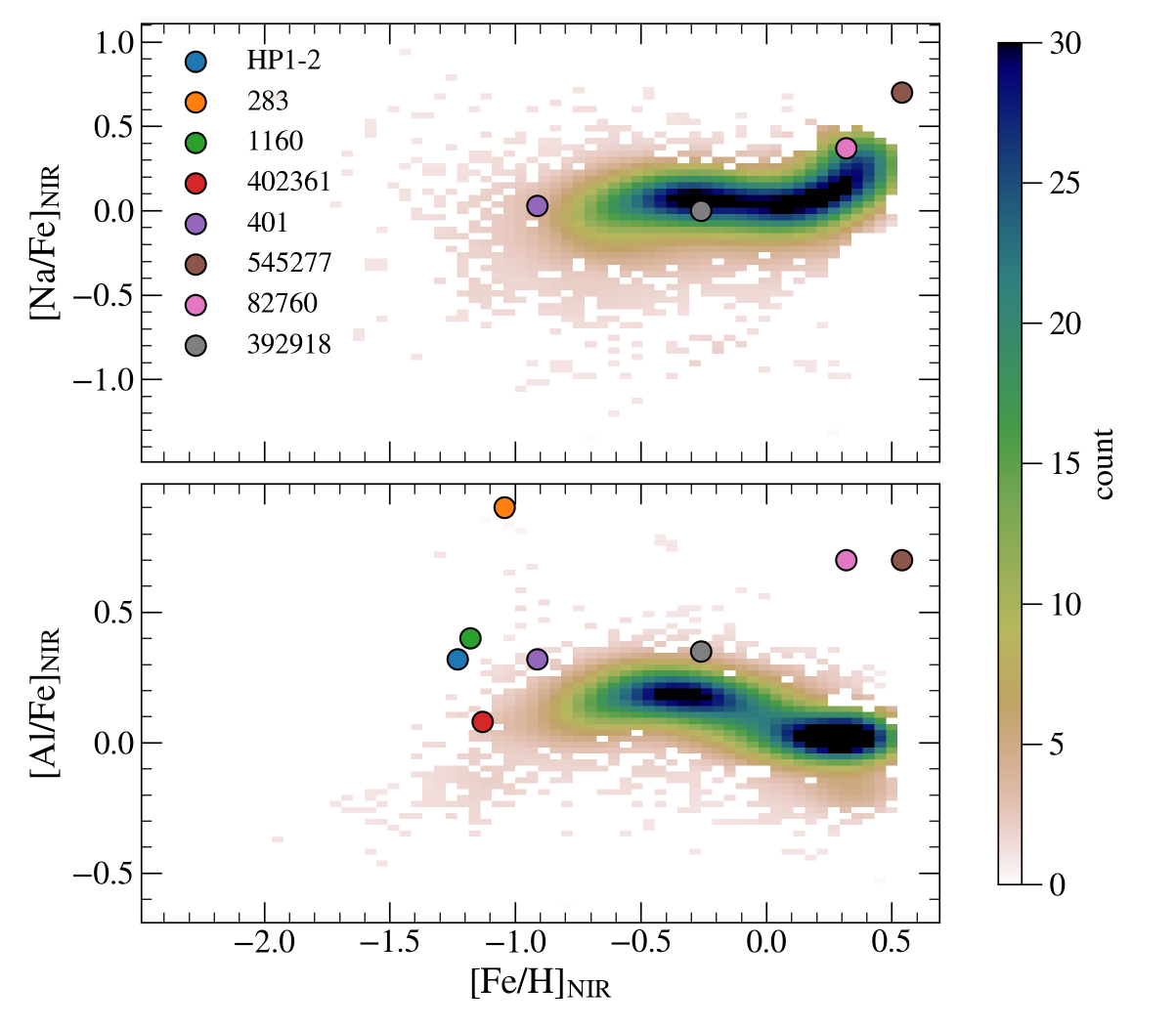}
    \caption{[Na/Fe] and [Al/Fe] vs. [Fe/H] for the final results given in Table 9, using the \textit{H}-band spectra,
    over-plotted on the bulge sample by \citet{queiroz21}.}
    \label{nirnaal}
\end{figure}

\begin{figure*}
\centering
\floatbox[{\capbeside\thisfloatsetup{capbesideposition={right,center},capbesidewidth=3.5cm}}]{figure}[\FBwidth]
{\caption{Same as Fig. \ref{nirnaal} but for [$\alpha$-elements/Fe]  vs. [Fe/H]. }
    \label{niralpha}}
{\includegraphics[width=5.5in]{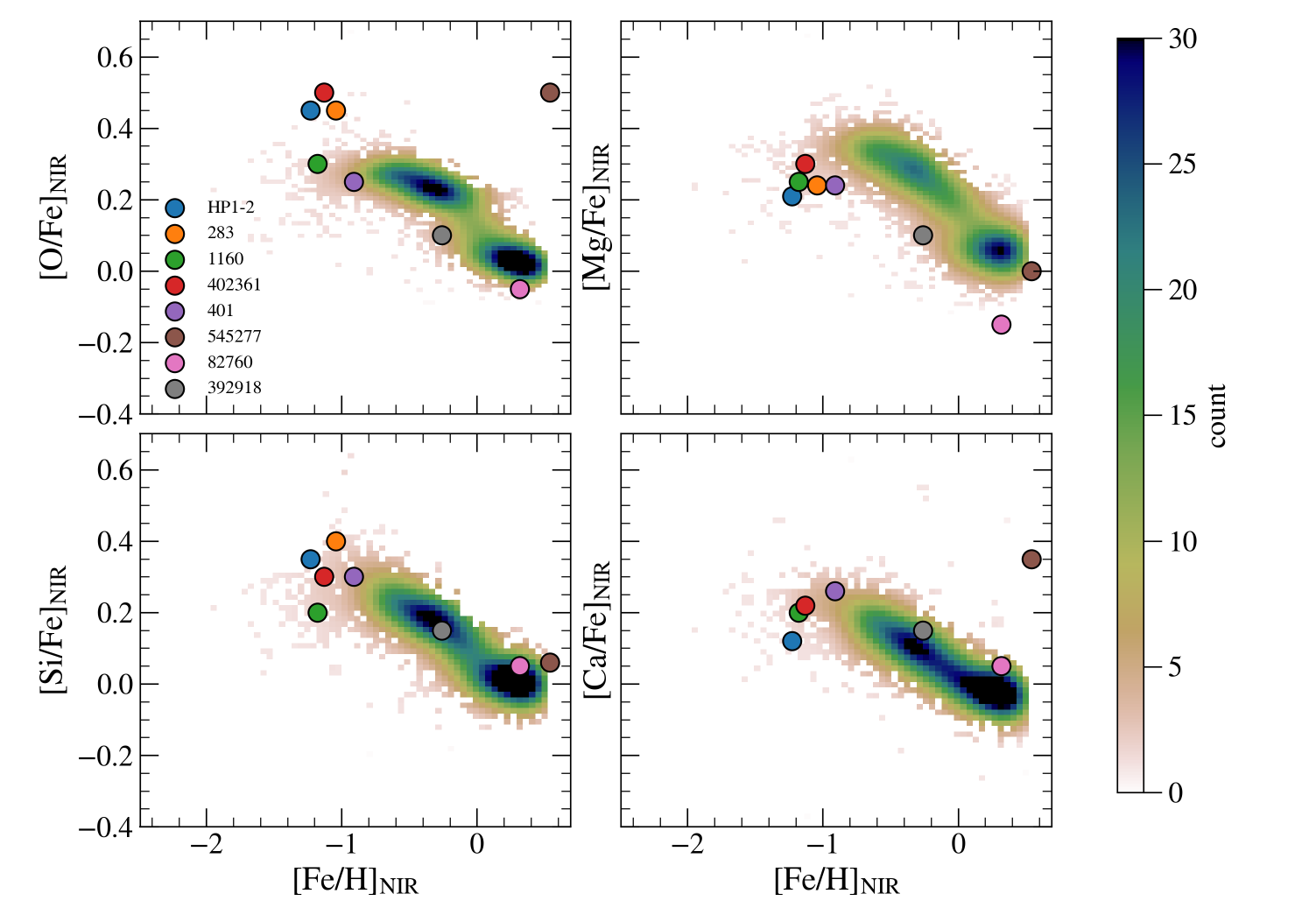}}
\end{figure*}

The sample is compared with the  bulge field stars  from the reduced proper motion (RPM) sample by \citet[hereafter Q21]{queiroz21}. Figures \ref{nirnaal} and \ref{niralpha} show our sample over-plotted on the Q21 sample. Our data are a good fit to the large sample by Q21, and we can say that, although our sample is small, it does represent the behaviour of bulge stars for the alpha elements and the odd-Z elements Na and Al.

\subsection{Errors}

The line-fitting procedure, given the high resolution of the spectra, has an error of $\sim\pm 0.1$ for
all lines. The error in abundances depends for the most part on the stellar parameters, with
the presence of blends being a secondary source of errors, as discussed for some of the lines (e.g., for \ion{Ca}{I} lines in the \textit{H}-band).

\begin{figure}
    \centering
    \includegraphics[width=8.5cm]{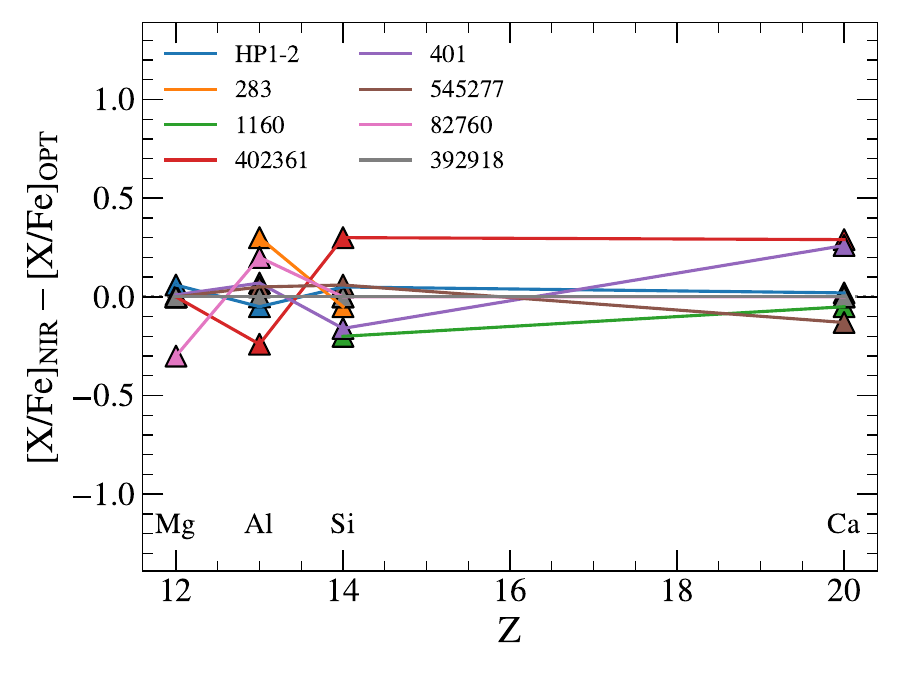}
    \includegraphics[width=8.5cm]{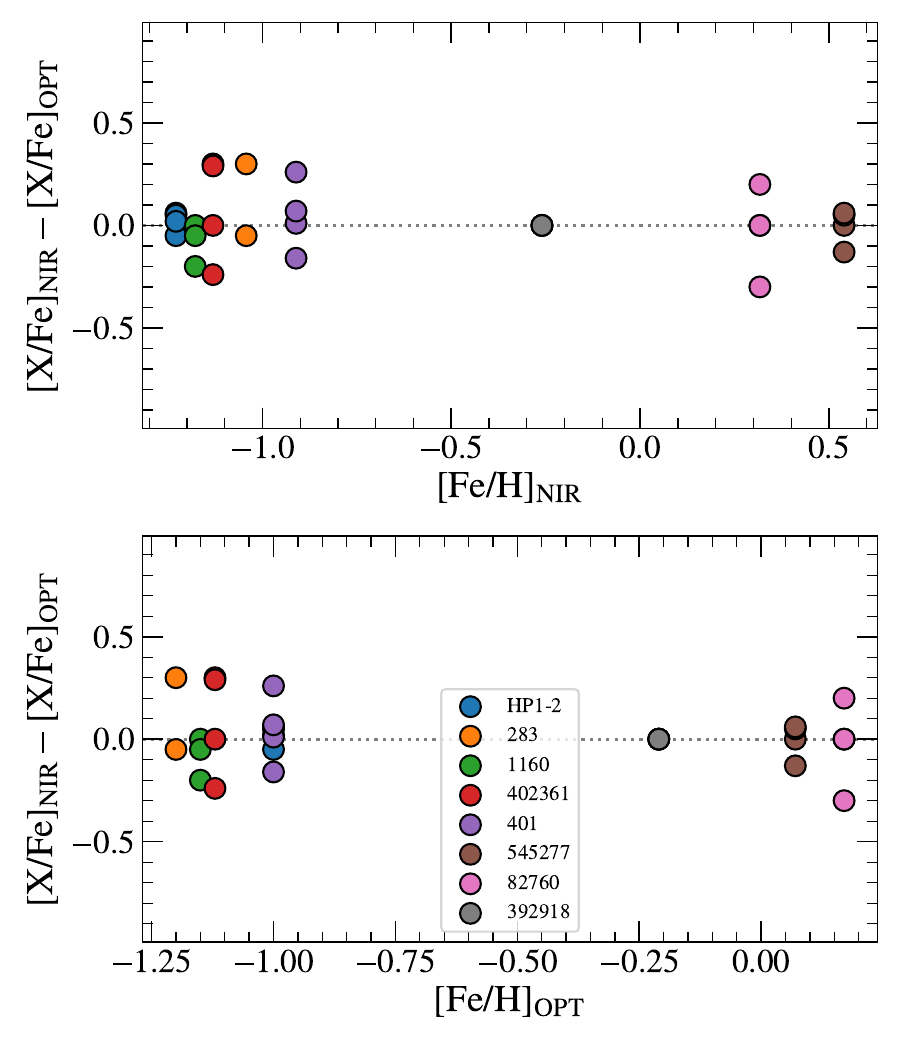}
    \caption{Abundance difference between \textit{H}-band and optical (Table 9) as a function of atomic number Z,
    and per star vs. [Fe/H]. }
    \label{compzfe}
\end{figure}

Figure \ref{compzfe} gives the abundance differences between the derivation in the \textit{H}-band
and optical based on the final stellar parameters (Table \ref{niropt}), both as a function of
atomic number Z and per star vs. [Fe/H].
The larger differences are for  Al ($\pm0.3$) and the smallest for Mg. These results show the good agreement between the derivation of abundances from the \textit{H}-band and optical, as long
as suitable stellar parameters are adopted.

\section{Conclusions}\label{sec6}

In this work, we compared abundance results obtained from optical spectra (VLT/UVES) and \textit{H}-band 
spectra (APOGEE) for eight bulge stars. The calculations were carried out adopting stellar parameters 
from the literature obtained from optical and \textit{H}-band spectra. We then computed abundances
employing the {\tt TURBOSPECTRUM} code throughout for the sake of homogeneity, adopting the two sets of stellar
parameters for the lines in the optical and the \textit{H}-band. The aim of this work is to find the best methods for deriving stellar parameters and their advantages and disadvantages, and to identify the
best lines for each of the analyzed elements.

A first and important conclusion is that the stellar parameters ---and in particular the effective temperatures obtained from ASPCAP--- that 
correspond to  a spectroscopic solution 
minimizing the {chi-squared differences between the observations and the models} in seven dimensions, lead to reliable parameters.
This is due to the derivation of C, N, and O abundances through
the strength of molecular lines of CO, CN, and OH, leading to reliable effective temperature, and more so when dealing with high-S/N spectra. 
This is an important compensation for the lack of \ion{Fe}{I} lines
of varied excitation energies, which are used to constrain effective temperatures in the optical.

In conclusion, as concerns stellar parameters,
previous APOGEE and UVES results are suitable for stars HP1-2, NGC6522-402361 and Palomar6-401.
A toohigh effective temperature was estimated for the stars N6558-283 and NGC6558-1160
from the optical, therefore the ASPCAP parameters are more suitable.
For the 3 metal-rich stars the optical parameters derived are more suitable than the ASPCAP ones.

Regarding the atomic lines, the Mg, Al, and Si lines in both \textit{H}-band and optical are suitable, 
with Si having stronger lines in the \textit{H}-band than in the optical.
For Ca, the optical appears to have a greater number of reliable lines. Na and Ti do not have
reliable lines in the \textit{H}-band, and therefore the optical lines are more suitable.

This findings of study will be particularly useful for future observations with the ELT/MOSAIC spectrograph, which will allow
simultaneous observations of the \textit{H}-band and selected regions in the optical.

Finally, a comparison of abundances in the presently studied 8 stars with bulge field stars by \citet{queiroz21} shows good agreement,and despite
the limited number of stars in our sample, they cover the entire  bulge metallicity range.

\section*{Acknowledgements}
PS acknowledges Funda\c{c}\~ao de Amparo \`a Pesquisa do Estado de S\~ao Paulo (FAPESP) post-doctoral fellowships 2020/13239-5 and 2022/14382-1.
BB acknowledges grants from FAPESP, Conselho Nacional de Desenvolvimento Científico e Tecnol\'ogico (CNPq) and Coordena\c{c}\~ao de Aperfei\c{c}oamento de Pessoal de N\'ivel Superior (CAPES) - Financial code 001. SOS acknowledges a FAPESP PhD fellowship no. 2018/22044-3.
PS, BB, HE, and SOS are part of the Brazilian Participation Group (BPG) in the Sloan Digital Sky Survey (SDSS), from the
Laborat\'orio Interinstitucional de e-Astronomia – LIneA, Brazil.
J.G.F-T gratefully acknowledges the grant support provided by Proyecto Fondecyt Iniciaci\'on No. 11220340, 
Proyecto Fondecyt Postdoc No. 3230001 (Sponsored by J.G.F-T)  and from the Joint Committee ESO-Government of Chile 2021 (ORP 023/2021), and 2023 (ORP 062/2023). D.G-D gratefully acknowledges support from the Joint Committee ESO-Government of Chile 2021 (ORP 023/2021), as well as the support from call N° 785 of 2017 of the Colombian Departamento Administrativo de Ciencia, Tecnología e Innovación (COLCIENCIAS).
Apogee project: Funding for the Sloan Digital Sky Survey IV has been provided by the Alfred P. Sloan Foundation, the U.S. Department of Energy Office of Science, and the Participating Institutions. SDSS acknowledges support and resources from the Center for High-Performance Computing at the University of Utah. The SDSS web site is www.sdss.org. 
SDSS is managed by the Astrophysical Research Consortium for the Participating Institutions of the SDSS Collaboration including the Brazilian Participation Group, the Carnegie Institution for Science, Carnegie Mellon University, Center for Astrophysics | Harvard \& Smithsonian (CfA), the Chilean Participation Group, the French Participation Group, Instituto de Astrofísica de Canarias, The Johns Hopkins University, Kavli Institute for the Physics and Mathematics of the Universe (IPMU) / University of Tokyo, the Korean Participation Group, Lawrence Berkeley National Laboratory, Leibniz Institut für Astrophysik Potsdam (AIP), Max-Planck-Institut für Astronomie (MPIA Heidelberg), Max-Planck-Institut für Astrophysik (MPA Garching), Max-Planck-Institut f\"ur Extraterrestrische Physik (MPE), National Astronomical Observatories of China, New Mexico State University, New York University, University of Notre Dame, Observatório Nacional / MCTI, The Ohio State University, Pennsylvania State University, Shanghai Astronomical Observatory, United Kingdom Participation Group, Universidad Nacional Autónoma de México, University of Arizona, University of Colorado Boulder, University of Oxford, University of Portsmouth, University of Utah, University of Virginia, University of Washington, University of Wisconsin, Vanderbilt University, and Yale University.

\section*{Data Availability}

 The spectra observed in the optical with the UVES spectrograph are available at the archival data from the European Southern Observatory (ESO).
 The spectra observed in the \textit{H}-band are from the APOGEE survey.



\end{document}